# Modeling of a Single Multimode Fiber Imaging System

**Chen Liu[1,2], Liang Deng[1,3], Deming Liu[2] and Lei Su[1,3,*]**

[1] Department of Electrical Engineering and Electronics, University of Liverpool, Liverpool L69 3GJ, UK
[2] School of Optical and Electronic Information, Huazhong University of Science and Technology, Wuhan 430074, China
[3] School of Engineering and Materials Science, Queen Mary University of London, London E1 4NS, UK
[*] l.su@qmul.ac.uk

**Abstract:** We present a detailed theoretical analysis on image transmission via a single multimode fiber (MMF). A single MMF imaging model is developed to study the light wave propagation from the light source to the camera, by using free-space Fourier optics theory and mode-coupling theory. A mathematical expression is obtained for the complete single MMF imaging system, which is further validated by image-transmission simulations. Our model is believed to be the first theoretical model to describe the complete MMF imaging system based on the transmission of individual modes. Therefore, this model is robust and capable of analyzing MMF image transmission under specific mode-coupling conditions. We use our model to study bending-induced image blur in single-MMF image transmission, and the result has found a good agreement with that of existing experimental studies. These should provide important insights into future MMF imaging system developments.

## 1. Introduction

The use of a tiny probe for image transmission has attracted much attention owing to its great potential in in-vivo endoscopy. To achieve this, a conventional approach is to reduce the size of the imaging fiber bundle, i.e., by reducing the number of individual fibers contained in the bundle. This, unavoidably, results in low-resolution images. Another technology uses a scanning single-fiber probe. In such an approach, a single-mode optical fiber was installed with a piezo-actuator at the distal end, such that the optical fiber tip can be scanned across the area of interest to acquire an image [1]. The use of mechanical actuators, however, increases the size of the imaging probe. A recent study showed that images can be reconstructed after transmitting through a single multimode fiber (MMF) via phase conjugation, leading to a scanner-free and wide-field single MMF imaging system [2]. Such a single MMF imaging probe offers many advantages compared to existing fiber bundles and scanning single-fiber probes, including ultra-high resolution close to the diffraction limit, and an ultra-thin diameter of a few hundreds of microns [1-3].

There have been considerable experimental investigations on the single MMF imaging system in recent years. In [4], special sampling patterns were used to enhance the image resolution and to reduce noise. In order to overcome the image distortion caused by fiber bending, a coherent beacon source was proposed to be placed at the distal tip of the MMF to obtain the bending information by comparing speckle patterns [5]. Additionally, digital phase conjugation was employed to reduce image distortion caused by modal scrambling and to generate focused spots through multimode fiber, thereby achieving high-resolution and lensless endoscopes [6]. This technique was further developed by adding a scattering medium in front of multimode fiber, leading to higher resolution and a longer working distance [7]. In [8], a single MMF confocal microscope was demonstrated, where collected signals were filtered with a virtual digital pinhole to obtain high-contrast images. Apart from the MMF confocal microscope, a scanning-free computational microscope was demonstrated to perform 3D fiber imaging. Calibration was performed by applying point sources at different locations. Single-frame image was then captured and reconstructed with a nonlinear optimization algorithm [9]. Very recently, it was shown in Ref [10] that even significant bent fibers can be predictable, thus allowing calculation of their transmission matrixes if the fiber shape is known. Single multimode fiber imaging with bending flexibility to some extent was also demonstrated [10]. In a recent theoretical work [11], a MMF design framework was presented, where the MMF coupling was modelled by Haar-distributed random unitary matrix without considering the transmission and coupling of individual modes.

The number of modes guided in an MMF usually exceeds tens of thousands, determined by factors such as the fiber diameter and wavelength. These spatial modes will unavoidably be coupled to each other during the transmission. Consequently, an image will be scrambled when being transmitted from the distal end to the proximal end of the MMF probe. In such an image transmission, every single pixel on the object will have a distinctive point spread function (PSF) according to the spatial mode coupling [2,3,12]. In order to reconstruct the original image from the scrambled output, one needs to find the relation between the input and output [12,13]. In such cases, transmission matrix is often used to represent the PSF from the input to the output [14,15]. Additionally, different from imaging system in transmission mode [14,15], optical-fiber-based imaging, such as endoscopy, usually operates in reflection mode. Therefore, the illumination process, from the light source via the imaging MMF to the object, must be investigated. By combining the speckle illumination and transmission matrix, the object image can then be reconstructed [2,16]. Therefore, it is crucial to develop MMF imaging theories and to understand and predict the imaging mechanism, thereby designing effective experimental approaches to address the existing main challenges such as the imaging MMF flexibility challenge (i.e., distortion caused by bending) and to enhance the system robustness and efficiency.

In this paper, we attempt to develop a theoretical model to describe the single MMF imaging system. Main problems being addressed include the propagation and diffraction of optical fields in free space and in the MMF. Our focus is centered on building a mathematical model and to explore the influence of the MMF on the imaging transmission. Our model is verified by conducting simulations for illumination and image reconstruction. Different from existing theoretical work on MMF imaging, our model is based on analyzing the propagation of individual modes, which is the first complete model describing the MMF imaging from the light source to the camera by studying the transmission of each guided modes. Therefore, our model has the capability to study specific mode coupling conditions in the MMF and provides insightful design guidelines for future flexible MMF imaging systems or endoscopes. As an application, we use our model to predict the effect of MMF bending on image transmission, which is an important design consideration in MMF imaging. This paper is organized as follows: We firstly discuss the complex field transmission in the optical system involving an MMF, using fiber mode-coupling theory and free-space diffraction theory. We then investigate the imaging conditions and validate the image recovery method. Numerical simulation is performed to study the propagating wave field on different observation planes and subsequently to verify the algorithm for image restoration. Finally, we use our model in combination with mode-coupling theory to investigate how bending-induced mode coupling changes the transmission matrix and influences the image recovery.

## 2. Theory

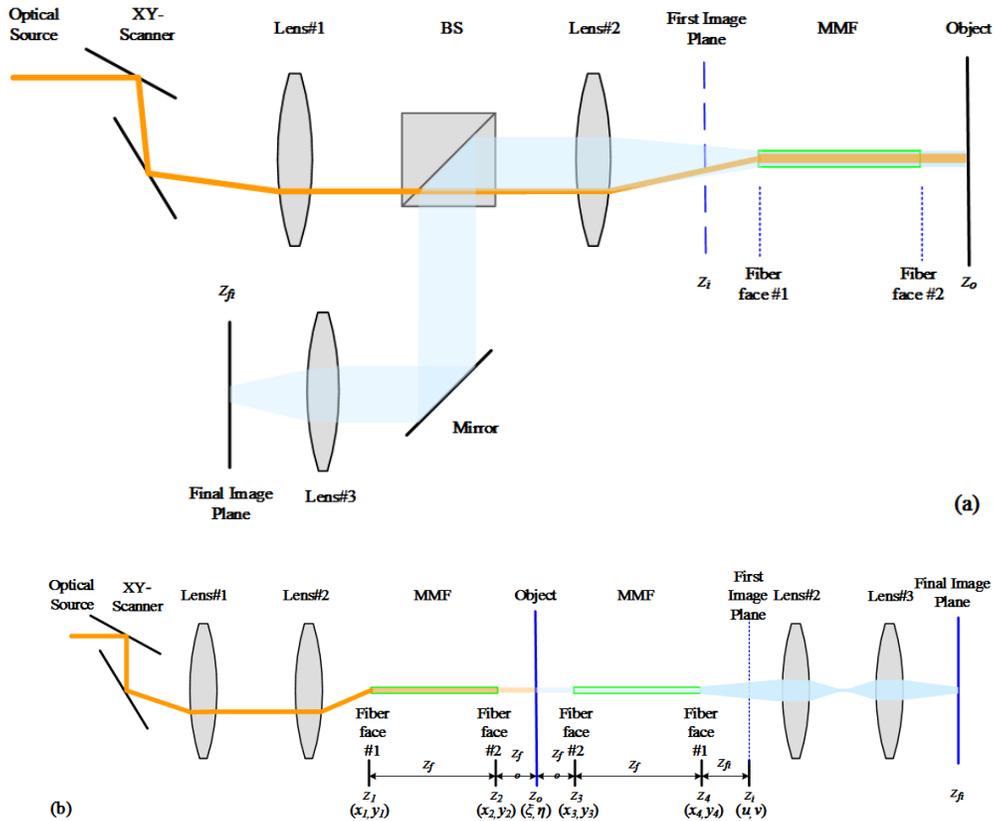

Fig. 1. (a) Single MMF imaging system model, and (b) the equivalent imaging system for (a).

A single-MMF image transmission system is shown in Fig.1. An MMF guides the image from the object plane ($z_0$) to the image plane ($z_i$). A monochromatic, linearly polarized Gaussian beam is relayed by lens #1, refocused by lens #2 and then coupled into the MMF (Fiber face #1), to illuminate the object at the distal end (Fiber face #2). Subsequently, the light reflected from the object, is coupled back into the MMF, and propagates along the fiber to the proximal end (Fiber face #1). Then the returned image at the first image plane ($z = z_i$) passes through the Lenses #2 and #3 to the final image plane ($z = z_{fi}$). The fiber radius and length are denoted by $a$ and $z_f$ respectively. The distance between the fiber and object is $z_{fo}$. For clarity, Cartesian coordinate system $(x_i, y_i, z_i)$ is used for the $i^{th}$ observation plane, and $(\xi, \eta)$ and $(u, v)$ coordinates are given for the object plane and image plane, respectively. The image transmission from $z_i$ to $z_{fi}$ can be regarded as ideal diffraction-limited optical transmission, which only produces an inverted magnified image. Therefore, the image transmission from $z_i$ to $z_{fi}$ is not taken into consideration in the following analysis.

In Fig. 1, the illuminating wave field can be described by a complex amplitude distribution [17]

$$u_s\left(x, y, z\right) = A_0 \frac{W_0}{W(z)} \exp\left[-\frac{\left(x^2 + y^2\right)}{W^2(z)}\right] \exp\left[-j\left(kz + k\frac{\left(x^2 + y^2\right)}{2R(z)} - \tan^{-1}\frac{z}{z_0}\right)\right] \qquad (1)$$

where k is the wavenumber ($k = 2\pi/\lambda$), $\lambda$ is wavelength of the incident light, $A_0$ is the peak amplitude, $W_0$ is beam radius, and $z_0$ is the Rayleigh length. $W(z)$ and $R(z)$ are the beam width and the wavefront-curvature radius, respectively. The mode field (electrical field) excited by the incident light can be written as a linear superposition of all guided modes in the imaging MMF [18]

$$u_1\left(x_1, y_1\right) = \sum_{m=1}^{M}\left[c_{m,in}E_m\left(x_1, y_1\right)\right] \qquad (2)$$

where $E_m$ is the normalized complex electric field amplitude distribution of $m^{th}$ propagation mode, and M is the total number of modes supported by the fiber. Here, we assume that there is no coupling between the radiation modes and guided modes. The excitation amplitude of the $m^{th}$ mode $c_{m,in}$ is governed by the complex overlap integral of the incident optical field and the complex conjugate of individual guide mode field [18]

$$c_{m,in} = \iint_{z=z_1}\left[u_s\left(x_1, y_1\right)E_m^*\left(x_1, y_1\right)\right]dx_1 dy_1 \qquad (3)$$

In Eq. (3), $E_m^*(x, y)$ is the conjugate complex electric field for the $m^{th}$ propagation mode in the fiber, assuming that the imaging MMF is a weakly-guiding fiber [19]. After propagating a distance of $z_f$ in the fiber, the field at the distal end (Fiber face #2) is given by

$$u_2\left(x_2, y_2\right) = \sum_{m=1}^{M}\left[c_{m,in}E_m\left(x_2, y_2\right)\exp\left(-j\beta_m z_f\right)\right] \qquad (4)$$

where $\beta_m$ is the propagation constant for the m$^{th}$ mode.

The propagation of the optical wave in the free space can be regarded as propagating in a diffracting system [20]. Thus Eq. (5) can be used to calculate the complex amplitude of the incident wave for object plane illumination:

$$u_o^-\left(\xi, \eta\right) = u_2\left(x_2, y_2\right) \otimes h_{2o}\left(\xi, \eta\right) \qquad (5)$$

where the impulse response from the plane z = $z_2$ to the object plane is[20]

$$h_{2o}\left(\xi, \eta\right) = \begin{cases} \dfrac{\exp\left\{jkz_{2o}\left[1+\left(\dfrac{\xi}{z_{2o}}\right)^2+\left(\dfrac{\eta}{z_{2o}}\right)^2\right]^{1/2}\right\}}{j\lambda z_{2o}\left[1+\left(\dfrac{\xi}{z_{2o}}\right)^2+\left(\dfrac{\eta}{z_{2o}}\right)^2\right]} (Rayleigh - Sommerfield) \\[20pt] \dfrac{\exp\left(jkz_{2o}\right)}{j\lambda z_{2o}}\exp\left[j\dfrac{k}{2z_{2o}}\left(\xi^2+\eta^2\right)\right] (Fresnel - Approximation) \end{cases} \qquad (6)$$

For Eq (6), the distance between the fiber endface and the object determines which approximation is applied in practice. The illumination light is then reflected by the object and the reflected field can be written as

$$u_O^+\left(\xi, \eta\right) = u_O^-\left(\xi, \eta\right)r\left(\xi, \eta\right) \qquad (7)$$

where $r(\xi, \eta)$ is the reflectance function of the object. When it reaches the fiber endface the reflected optical wave has a field distribution:

$$u_3^-(x_3, y_3) = u_O^+(\xi, \eta) \otimes h_{o3}(x_3, y_3) = r(\xi, \eta)[u_2(x_2, y_2) \otimes h_{2o}(\xi, \eta; x_2, y_2)] \otimes h_{o3}(x_3, y_3; \xi, \eta) \tag{8}$$

Then the light field coupled into fiber again, and similarly, the electric field of guided optical wave coupled into the MMF can be decomposed into a series of guided modes:

$$u_3^+(x_3, y_3) = \sum_{m=1}^{M} \left[ c_{m,out} E_m(x_3, y_3) \right] \tag{9}$$

The parameter $c_{m,out}$ denotes the mode m coupling coefficient, and is given by

$$C_{m,out} = \iint_{z=z_3} \left[ u_3^-(x_3, y_3) E_m^*(x_3, y_3) \right] dx_3 dy_3 \tag{10}$$

The wave field leaving the optical fiber can then be described by

$$u_4(x_4, y_4) = \sum_{m=1}^{M} \left[ c_{m,out} E_m(x_4, y_4) \exp\left(-j\beta_m z_f\right) \right] \tag{11}$$

Finally, the output image in the image plane can be obtained by calculating the convolution of the impulse response

$$u_I(u, v) = u_4(x_4, y_4) \otimes h_{4i}(u, v; x_4, y_4) \tag{12}$$

It can be seen from the above derivation, that the impulse response from the object to the image plane is a complex space variant function, which is related to illuminating conditions. This makes the recovery of the original image extremely difficult. If we make $[u_2(x_2, y_2) \otimes h_{2o}(\xi, \eta)]$ a constant (this is true when the object plane is illuminated uniformly), the complex transfer function of optical field from the light source to the object (via the MMF) can be neglected. In this way, the imaging process is simplified to the transmission of the image from the distal end to the proximal end of the MMF. Additionally, in MMF imaging, the distal end of the MMF is made very close to the object (i.e., $z_{fo}$ and $z_{fi}$ are very small) and the diameter of the MMF is in the order or hundreds of micrometers. Therefore, Fresnel condition is satisfied and Fresnel approximation can be used in Eq (6) to describe the optical wave propagation in free space. For impulse response calculation, the object is considered as a $\delta$ function at point $(\xi, \eta)$ in the object plane. From Eq. (6), the impulse response function from the point $p_o(\xi, \eta)$ on object plane to the point $p_3(x_3, y_3)$ on fiber face can be written as

$$u_{o3}(x_3, y_3; \xi, \eta) = r(\xi, \eta) h_{o3}(x_3, y_3; \xi, \eta)$$

$$= r(\xi, \eta) \frac{\exp\left(jkz_{fo}\right)}{j\lambda z_{fo}} \exp\left\{ j\frac{k}{2z_{fo}} \left[ (x_3 - \xi)^2 + (y_3 - \eta)^2 \right] \right\} \tag{13}$$

By substituting Eq. (13) into Eq.(10) it yields the excited modal amplitude

$$c_{m,out}(\xi, \eta) = r(\xi, \eta) \frac{\exp\left(jkz_{fo}\right)}{j\lambda z_{fo}} \iint_{z=z_3} \exp\left\{ j\frac{k}{2z_{fo}} \left[ (x_3 - \xi)^2 + (y_3 - \eta)^2 \right] \right\} E_m^*(x_3, y_3) dx_3 dy_3$$

$$= r(\xi, \eta) a_m(\xi, \eta)$$

$$\tag{14}$$

where $a_m(\xi, \eta)$ is the coupling coefficient for fiber mode m when r =1, i.e. the unit impulse response. Similarly, based on Eq. (6), the impulse response function from the fiber endface to the image plane can be written as

$$h_{4i}(u, v; x_4, y_4) = \frac{\exp\left(jkz_{fi}\right)}{j\lambda z_{fi}} \exp\left\{ j\frac{k}{2z_{fi}} \left[ (u - x_4)^2 + (v - y_4)^2 \right] \right\} \tag{15}$$

Finally, by substituting Eqs. (13-15) into Eqs. (11-12), we obtain the response of the object pixel $p_o(\xi, \eta)$ on the image plane:

$$u_{oi}(u,v;\xi,\eta) = r(\xi,\eta)\frac{exp(jkz_{fi})}{j\lambda z_{fi}}\sum_{m=1}^{M}\iint a_m(\xi,\eta)\exp(-j\beta_m z_f)$$

$$exp\left[j\frac{k}{2z_{fi}}\left[(u-x_4)^2+(v-y_4)^2\right]\right]E_m(x_4,y_4)dx_4dy_4$$

(16)

Eq. (16) is the complete expression for the image transmission in the single MMF imaging system. This expression shows the impulse response of a linear space-variant system. The object and the image are correlated by an overlapping integral instead of a convolution integral. This space-variant system is a result of the optical-field transmission through the MMF.

Since both the object and the image may consist of many individual pixel points, these pixels can be taken row by row and stored in a column in a matrix, where their location coordinates were denoted by $n_o$(for the object) and $n_i$ (for the image), respectively. If N is the total number of the pixels of an image, $u_{oi}(n_i,n_o)$ can be rewritten to represent the transmission of individual pixels.

$$u_{oi}(n_i,n_o)=h_{oi}(n_i,n_o)r(n_o)=r(n_o)\frac{\exp(jkz_{fi})}{j\lambda z_{fi}}\sum_{m=1}^{M}\left[\sum_{n_4=1}^{N}\left[t_{fi}(n_i,n_4)f(n_4,m)c(m,n_o)\right]\right]$$

(17)

where, $h_{oi}$ can be regarded as the PSF of this image system, and is determined by individual wave transmission process. $c(n_o,m)$ represents the change of the guided fiber mode $m$ excited by object pixel $n_0$, influenced by mode coupling, mode dispersion, and mode excitation from the object to imaging fiber. $t_{fi}(n_i,n_4)$ denotes the transmission coefficient from a point $n_4$ in fiber proximal end ($z = z_4$) to a point $n_i$ in the image plane. $f(n_4,m)$ is the value of the complex electrical field of fiber mode $m$ at point $n_4$. The matrix forms for the above are $C$, $T_{fi}$, and $F$, respectively. $r(n_o)$ is the complex reflectance of the point $n_o$ on the object plane, which forms a diagonal matrix $R$.

Eq. (17) suggests that the object can be regarded as a set of point sources, and each point source produces its own image at the image plane. The final image is then the superposition of the impulse responses from all these point sources. Thus, Eq. (17) can be simplified as

$$U_i=\left(T_{fi}FC\right)R=T_bR$$

(18)

where $U_i$ is the output image in the image plane when uniform illumination is achieved. $T_b$ is the transmission matrix from the object plane to the image plane. In the experiment study in [3], $T_b$ was obtained through calibration, by placing a point light source at the object plane without and with the imaging MMF, to obtain the field distribution $S$ and $U_{i\_s}$, respectively. Then $T_b$ is calculated by using $T_b=U_{i\_s}S^{-1}$. With both the magnitude and phase information of the input and output , the image of the object can be restored:

$$R=\left(T_{fi}FC\right)^{-1}U_i=T_b^{-1}U_i$$

(19)

In summary, based on the above analysis, the image reconstruction in a single MMF imaging system is described as follows. The first step is to calibrate the transmission matrix. On approach is to alter the scanning angle $\theta_i$ of the light source, and record its image on the image plane as shown in Fig.1. Subsequently, the image can be transformed into a one-dimensional row array $U_{\theta_i}=[U_{n1}\cdots\;U_{ni}\;\;\;U_{nN}]_{\theta_i}$. Then the image matrix including different launching conditions, $U_{cal}$, can be constructed as $U_{cal}=\left[U_{\theta_1}^T\;U_{\theta_2}^T\;\cdots\;U_i^T\;\cdots\;U_{\theta_N}^T\right]$ . Similarly, image matrix of input light source including different launching conditions, $R_{cal}$, can be obtained without the MMF: $R_{cal}=\left[R_{\theta_1}^T\;R_{\theta_2}^T\;\cdots\;R_i^T\;\cdots\;R_{\theta_N}^T\right]$ . Then the transmission matrix can then be calculated using

$$T_b=U_{cal}R_{cal}^{-1}$$

(20)

Once transmission matrix is obtained, the image can then be successfully recovered using Eq. (19).

In our model, the radiative modes and cladding modes are not considered. This is because the coupling between guided modes and radiative or cladding modes is negligible when the fiber is straight or only bent slightly. For tighter bends, the coupling to the radiative or cladding modes mainly leads to higher transmission loss. Compared to guided-mode coupling, bending-induced loss has negligible impact on image transmission via a short MMF. The polarization mixing in a short MMF is relatively weak compared with spatial mode coupling [10]. Therefore, in our model we only consider spatial mode coupling and the polarization mode mixing is neglected.

## 3. Results and discussion

It can be seen from Eq. (19), that both matrix $T_{fi}$ and mode field distribution $F$ are constant once the imaging system is established. The only variable is matrix $C$, which varies as a result of the mode coupling, e.g., when the imaging MMF is moved or bent [21]. By monitoring $C$, the transmission matrix $T_b$ can be evaluated in real time [5,22], thereby tackling the flexibility challenge in single MMF imaging.

To achieve uniform illumination, a simple approach could be the altering of the launching conditions of the illumination laser source, for example, to tilt or offset the incident laser beam at the MMF input. Such a launching condition excites most of the guided fiber modes. The mode power distribution is determined by the overlapping integral as shown in Eq. (14). With sufficient fiber modes excited, the overall electric field can be regarded as uniform. This method was applied to mimic an image illuminated by using the speckle imaging method to eliminate the on-the-way-in distortion [2], and then the speckle-free image was obtained by the random laser illumination method [23].

In our simulation, the incident beams are tilted in both x and y directions with angles $\theta_x$ and $\theta_y$ respectively. This resembles the real experimental condition using an x-y scanning mirror. The incident Gaussian beam field distribution is given by

$$G = \begin{bmatrix} G_{\theta_1}^T G_{\theta_2}^T \text{ L } & G_{\theta_i}^T \text{ L } & G_{\theta_N}^T \end{bmatrix}^T \tag{21}$$

where the $i^{th}$ column denotes the field distribution for $i^{th}$ beam launching condition. The modal field distribution at the excitation end of the MMF is given by

$$F = \begin{bmatrix} F_{M_1}^T F_{M_2}^T \text{ L } & F_{M_i}^T \text{ L } & F_{M_M}^T \end{bmatrix}^T \tag{22}$$

where the $i^{th}$ column represents the field distribution of the $i^{th}$ fiber mode.

The mode coupling coefficients for all launching conditions can then be expressed by $C_{mi} = F^{*T} G$, in which the $(i, j)^{th}$ entry denotes the excited field amplitude of the $i^{th}$ fiber mode for the $j^{th}$ source launching condition. Then, the field distribution on fiber face #2 is given by $U_2 = F[D(F^{*T}G)]$. $D$ is a diagonal matrix containing the mode propagation constant $\beta_i$ and fiber length, where the element in the $i^{th}$ row is $exp\,(-j\beta_i z_f)$. Subsequently, the illumination field on the object plane can be written as

$$U_o^- = T_{fo}F[D(F^{*T}G)] = (T_{fo}FDF^{*T})G = T_f G \tag{23}$$

where the matrix $T_{fo}$ describes the input-output response from fiber face #2 to the object. Note that the response from object plane to fiber endface, $T_{of}$, is the transpose of $T_{fo}$. Similarly, the matrix $T_{fi}$ is the input-output response from fiber endface to the image plane.

Under different launching conditions, the complex field distribution on object plane can be expressed as $U_o^+ = RU_o^-$. Then, the field distribution reflected back into the fiber is $U_3 = T_{of}U_o^+$. The mode coupling coefficients in the reverse transmission can be expressed as $C_{mo} = F^{*T} U_3$. The field distribution on fiber face #4 is given by $U_4 = F[D(F^{*T}U_3)]$. The image field on the image plane is

$$U_i = T_{fi}F\left[D\left(F^{*T}T_{of}U_o^+\right)\right] = \left(T_{fi}FDF^{*T}T_{of}\right)\left(RU_o^-\right) = T_b U_o^+ \tag{24}$$

where $DF^{*T}T_{of}$ is the realization of matrix C in former theoretical analysis. if illumination is uniform, $\sum_{\theta_1}^{\theta_N} U_o^-(n_o, \theta_i) = constant$.

The MMF in the simulation is a 1m-long step-index fiber with a core diameter of 200 μm. The core and cladding refractive indices are 1.456 and 1.482, respectively. The distance between the fiber distal end and the object $z_{fo}$ is 1mm. The sampling pitch (resolution) is set to 2 $\mu m$. There are in total 9433 Linear Polarized (LP) modes guided in the fiber. We assume that the electric field of the Gaussian beam is linear polarized along the $x$-axis. The excitation source is a monochromatic Gaussian beam at 633 nm. The beam waist radius is set to 5μm and the spot is scanned on the object plane with the help of two axis scanner. To produce uniform illumination on the object plane, the source is scanned across the fiber proximal end. Meanwhile, the laser beam only offsets a small angle to the axis of the fiber, so that the weakly waveguide approximation condition is satisfied.

The simulation results of electric field information for resolving transmission matrix is illustrated in Fig. 2. It shows that the input Gaussian spots (Fig. 2(a) and (b)) are distorted by the fiber. The wave electric field after propagation in the fiber is scrambled as is shown in Fig. 2 (d) and (e). Fig.2 (c) and (f) are the amplitude and phase pattern of transmission matrix, which is obtained from the data in (a-b) and (d-e) using Eq. (24).

Fig. 3 shows how the original image is reconstructed in the single MMF imaging system. The original object pattern Fig. 3 (f) can be partially restored. However, distortions are introduced in the restored images. The distortion can be reduced by applying a more uniform illumination on the object plane.

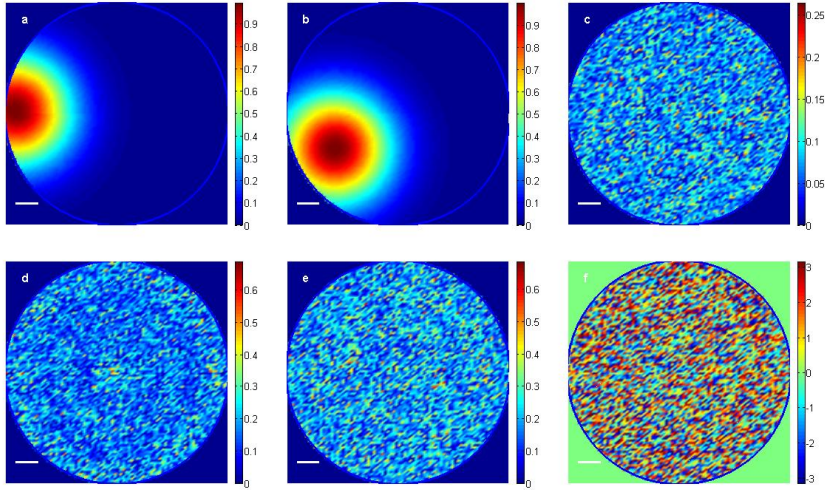

Fig. 2. Calculated electric field distribution in the MMF imaging system: Excitation field at MMF input with the excitation angle (a) $\theta_1$ and (b) $\theta_2$; (d) output field amplitude at the image plane with the excitation angle $\theta_1$ and (e) $\theta_2$. (c) amplitude and (f) phase of one column of the resolved transmission matrix. Scale bar: 20 μm; color bars are arbitrary units. The beam waist radius is set to 50 μm in our simulation in order to produce uniform illumination.

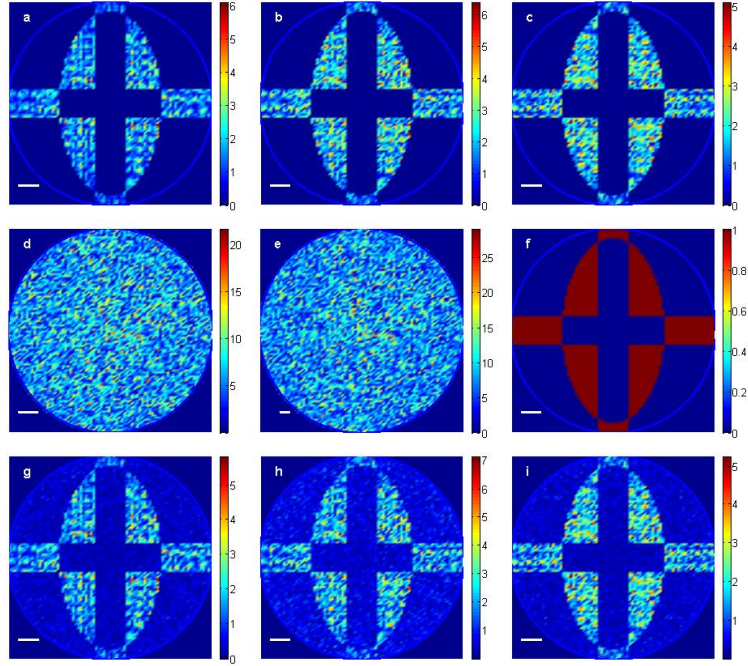

Fig. 3. The image reconstruction simulation results for the single MMF imaging system. The amplitude distribution on the object plane at two different excitation angles (a) $\theta_1$ and (b) $\theta_2$. (c) Averaged amplitude distribution on the object plane. The amplitude pattern at incident angles (d) $\theta_1$ and (e) $\theta_2$ on the image plane. The restored amplitude image at incident angles (g) $\theta_1$ and (h) $\theta_2$. (f) The object. (i)The reconstructed image averaged over all scanning angle. Scale bar: 20 $\mu m$, and colorbars are arbitrary units.

In the discussions below, we use our imaging model to study how the bending-induced mode coupling changes the transmission matrix and influences image transmissions via the imaging MMF. We know from the derivation in the previous section that in Eq. (19), any change in the mode coupling will lead to a new $C$ so as to change the transmission matrix. Such mode coupling can be expressed as [24]

$$\frac{dA_m}{dz} = \sum c_{nm} A_n e^{j(\beta_n - \beta_m)z} \qquad (25)$$

where $c_{nm}$ is the coupling coefficient between the $n$-th mode and the $m$-th mode and $A_n$ stands for complex amplitude of $n$-th mode. Eq. (25) suggests that the complex-amplitude changing rate $dA_m/dz$ of one mode is determined by the coupling with all other modes.

For a flexible single MMF imaging probe, any bending induced on the MMF leads to a new transmission matrix. The coupling coefficient $c_{nm}$ as a result of bending is given below [24]

$$c_{nm} = -j\frac{k^2}{2\beta_n}\iint n_d(r,\phi)E_n E_m r dr d\phi \qquad (26)$$

where , , $n_d$ stands for bending-induced index change and is expressed as [24]

$$n_d(r,\phi) = 2n_{core}^2 \cdot r \cdot \sin\phi \, / \, R \qquad (27)$$

where $n_{core}$ is initial fiber core index and $R$ stands for bending radius.

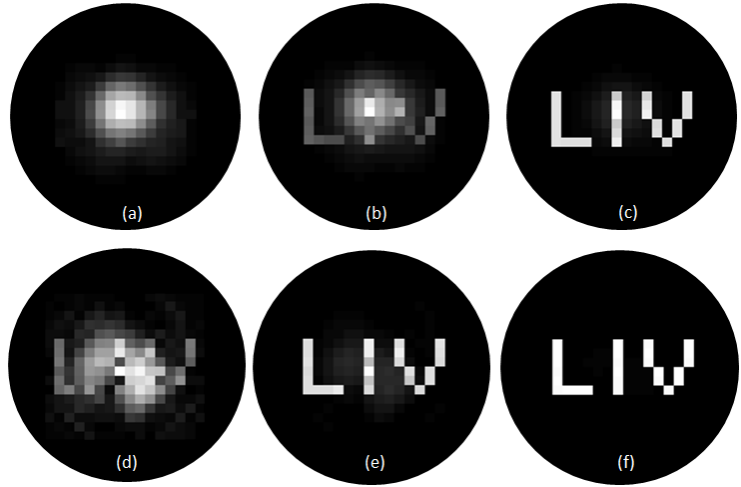

Fig. 4. Recovered images for bent fibers using the TMs of corresponding straight fibers. (a) 413 modes, R=1m; (b) 413 modes, R=10m; (c) 413 modes, R=100m; (e) 492 modes, R=1m; (e) 492 modes, R=10m; and (f) 492 modes, R=100m;

Fig. 4 shows the impact of bending on image recovery. Step-index MMFs at two different diameters (40um and 44um) are used in our simulation to study the influence of bending on image transmission. The number of guided modes supported by these two MMFs is 413 and 492, respectively. The length of the fiber in simulation is 1cm. The core and cladding refractive indices are 1.45 and 1.44, respectively. Without loss of the generosity, in order to reduce the simulation time, we use a shorter MMF with fewer modes compared to the MMF in the previous calculations. As illustrated, we attempt to recover images at different bending radius with the transmission matrix obtained for a straight fiber. In general, a tighter bend (a smaller bending radius) leads to a worse result in image recovery. In the meantime, interestingly, certain extent of bending tolerance can be observed, that is, a blurred but recognizable image can be obtained with the original transmission matrix at large bending radii. Similar bending-tolerance results were demonstrated experimentally in [2], and our analysis serves as a theoretical explanation to their experimental observations. In the case of a 1-m bending radius, the original object can hardly be seen even with fibers supporting 492 modes. This is because a tighter bend leads to larger mode coupling coefficients (i.e., stronger mode coupling). Therefore, severer image distortion is expected. In addition, we discovered that the number of modes also affects the image recovery. In Fig. 4, the 492-mode MMF exhibits a better tolerance in bending than that of the 413-mode MMF. This is because, we believe, the effect of bending is shared among all the modes prorogating in the MMF. By involving a larger number of modes, the effect resulted from the same bending becomes smaller for individual mode coupling coefficients, thereby resulting in a less blurred output image. Commercial MMFs usually support tens of thousands spatial modes, and therefore offers some degree of flexibility in image recovery.

## 5. Summary and conclusion

We presented the first complete theoretical model for a single MMF imaging system from the light source to camera, by evaluating the transmission and coupling of individual modes. Mathematical formulas were obtained to describe the imaging process. We discussed the importance of uniformed illumination and demonstrated the imaging reconstruction method based on our theory. Theoretical analysis showed that the image on the object plane can be reconstructed using a pre-determined transmission matrix. We also showed that the transmission matrix can be acquired through a calibration procedure by scanning a small radius beam on the object plane. Our work presented an insightful model for future single MMF imaging studies. For example, our model was used to predict the influence of bending-induced mode coupling in MMF imaging, and this should provide useful guidelines in designing future flexible single MMF image system.

**Acknowledgements**

This work was supported by the Engineering and Physical Sciences Research Council [grant number EP/L022559/1]; the Royal Society [grant number RG130230]; Chen Liu is grateful to China Scholarship Council for providing financial support; Liang Deng is grateful to the University of Liverpool for PhD funding.